\documentstyle[aps,preprint]{revtex}

\tightenlines

\begin{document}

\title{Microwave detected, microwave-optical double
resonance of NH$_{3}$,
NH$_{2}$D, NHD$_{2}$, and ND$_{3}$:  \\
II. Predissociation dynamics of the \~A state.\thanks{This
work is taken in part from the
Ph. D. thesis of Steven A. Henck.\protect\cite{Henck:1990}}}

\author{Steven A. Henck,\thanks{Present address: Texas Instruments, 13536 N.
Central Expy.%
, MS 992, Dallas, TX 75243}\   Martin A. Mason, Wen - Bin Yan,\thanks{Present
 address: Energia, Inc., P. O. Box 1468, Princeton, NJ 08542}
and Kevin K. Lehmann}
\address{Department of Chemistry, Princeton University,
Princeton, NJ 08544}

\author{and Stephen L.\ Coy}
\address{Harrison Regional Spectroscopy Laboratory,
Massachusetts Institute of Technology, Cambridge, MA 02139}

\maketitle

\draft %makes pacs numbers print

\begin{abstract}

	Using microwave detected, microwave-optical double resonance, we
have measured the homogeneous linewidths of individual rovibrational
transitions in the \~A state of NH$_{3}$, NH$_{2}$D, NHD$_{2}$, and ND$_{3}$.
We
have used this excited state spectroscopic data to characterize the height of
the
dissociation barrier and the mechanisms by which the molecule uses its
excess vibrational and rotational energies to help overcome this barrier. To
interpret the observed vibronic widths, a one dimensional, local mode
potential has been developed along a N-H(D) bond.  These calculations suggest
the
barrier height is roughly 2100~cm$^{-1}$, approximately 1000~cm$^{-1}$
below the {\it ab initio\/} prediction.
The observed vibronic dependence of levels containing two or more quanta
in $\nu_{2}$ is explained by a Fermi resonance between 2$\nu_{2}$ and the
N-H(D) stretch.
This interaction also explains the observed trends due to isotopic
substitution.
The rotational enhancement of the predissociation rates in the NH$_{3}$ 2$^{1}$
level
is dominated by Coriolis coupling while for the same level in ND$_{3}$,
centrifugal effects dominate.

\end{abstract}


\clearpage

\section{Introduction}

Although the electronic excited states of most molecules are known to
predissociate, the exact mechanisms of these processes remain largely
undetermined.
The difficulty in modeling molecular predissociation is that
the predissociation is usually caused by a coupling of a bound excited
electronic state to another electronic state in which the nuclear motion is not
bound\protect\cite{Herzberg:1966}. The rate of dissociation depends upon the
properties of both
electronic states, as well as the nonadiabatic coupling between them.
In contrast, the rate of predissociation of the ammonia \~A state
is determined by the rate of crossing over, or tunneling through, a small
barrier on a single potential energy surface\protect\cite{Runau:1977}.  While a
conical
intersection exists well beyond this barrier, this intersection effects only
the
product state distribution and not the predissociation rate.  Since knowledge
of only a single electronic surface is needed to understand the
\~A state predissociation, it represents one of the most
theoretically tractable chemical reactions.

The nature of the barrier to \~A state dissociation is well understood.  In
an early {\it ab initio} SCF and CI study, the \~A state was shown to result
from
excitation to the 4a$_{1}^{'}$ molecular orbital (MO) which continuously
changed from
a predominantly 3s Rydberg orbital at the equilibrium conformation to a
hydrogen 1s atomic orbital at large H$_{2}$N--H
separation\protect\cite{Runau:1977}.   A
small barrier exists along this coordinate due to the change from Rydberg to
 anti-valence character of this orbital as an N--H bond was stretched.  The
magnitude of this barrier was also shown to be a minimum for planar
geometry and to increase with out-of-plane deformation.  The minimum
barrier height remains a controversial quantity.  In the initial {\it ab
initio}
study by Runau {\it et al.}\protect\cite{Runau:1977}, the
barrier was predicted to lie 4600~cm$^{-1}$ above the \~A state minimum while a
more involved theoretical calculation that allowed the H--N--H angle to relax
has predicted this barrier to lie at 3200~cm$^{-1}$\protect\cite{McCarty:1987}.
Semi-classical
trajectories predict it lower by an additional
1200~cm$^{-1}$\protect\cite{Ashfold:1986a}.   An
accurate determination of this barrier height is essential to correctly
understand the predissociation mechanisms.  Hence, its determination was a
primary goal of the present work.

Besides the exhaustive theoretical
effort\protect\cite{Runau:1977,McCarty:1987,%
Ashfold:1986a,Dixon:1988,Rosmus:1987,Tang:1988,Tang:1990,Tang:1991},
an extensive experimental effort has also been
directed toward characterizing the barrier height and predissociation
mechanisms\protect\cite{Ashfold:1986a,Ashfold:1984,Ashfold:1985,Ashfold:1990,%
Biesner:1988,Biesner:1989,Douglas:1963,Endo:1990,Fuke:1988,Liefson:1933,%
Vaida:1987,Walsh:1961,Xie:1986,Ziegler:1984}.
Due to the lack of rotational data as discussed in the previous paper, much of
the current
understanding of the predissociation mechanism(s) has been extracted by
interpreting the vibronic dependence of the predissociation lifetimes.  In
particular, Vaida and coworkers\protect\cite{Vaida:1987}  observed a single
fourteen member progression in the absorption spectrum of NH$_{3}$ and ND$_{3}$
in a seeded supersonic jet.  Inhomogenous broadening was sufficiently
reduced such that for each vibronic member of this progression, an average
rovibrational predissociation rate was extracted.  In agreement with the
{\it ab initio\/} predictions\protect\cite{Runau:1977}, for both isotopomers,
the predissociation rates were
observed to be slowest for the 2$^{1}$ level due to the strong dependence of
the
barrier height on out-of-plane deformation.  Above this level for both species,
the predissociation rate was observed be enhanced greatly by increasing
$\nu_{2}$
excitation.  They postulated that for these higher lying levels anharmonic
interactions were coupling energy from the bound bending coordinate to the
unbound dissociation coordinate, thus,
circumventing the barrier.  Due to the strong anharmonicity of the \~A state
potential, only a limited amount of information about the barrier is likely to
be revealed through the study of these higher lying levels.

Deuterium substitution has been shown to reduce the predissociation
rate in \~A state ammonia substantially and particularly in the lowest few
vibronic levels\protect\cite{Ashfold:1985,Douglas:1963,Vaida:1987}.
Thus, determining predissociation lifetimes as a function of the number
of hydrogens substituted should be a sensitive measure of the barrier height.
This effect is attributed to the  slower tunneling rates of the deuterium
compared to hydrogen as well as slower tunneling rates due to the lower zero
point energy of the N-D stretching modes\protect\cite{Nakajima:1991}.  If one
neglects
the interaction among the three equivalent N-H(D) bands, the dissociation
rate would be the sum of tunneling rates through the three equivalent
barriers and each subsequent deuterium substitution would result in an
approximate constant decrease in predissociation rate.  Such an effect has
been observed in the \~A state action spectra from photodissociation through
the 2$^{1}$ level\protect\cite{Nakajima:1991}.    At even greater excitation,
coupling of
vibrational energy into the dissociation coordinate would be expected to
enhance the dissociation rate, since it provides a mechanism to `go over the
top' of the barrier.  Such resonances will depend sensitively on the
frequencies of the individual modes involved such that a simple isotopic
dependence would not be expected.  The effect of any vibration-rotation
interactions on the dissociation rate may also lead to a more complex isotopic
dependence.

Since the lowest two vibronic levels could only predissociate by
tunneling through this barrier, it was anticipated that a better picture of the
barrier would be revealed through an understanding of the predissociation in
the lowest few vibronic levels.  Hence, these levels have been probed by a
host of experimental techniques including
absorption\protect\cite{Douglas:1963},
two photon fluorescence excitation (2PFE)\protect\cite{Ashfold:1985},  and
rotational resonance Raman excitation profiles
(RRREP)\protect\cite{Ziegler:1987}. This
work has established a weak predissociation dependence on rotational level.
This rotational dependence has been quantitatively fit to a model incorporating
a centrifugal modification to the barrier
height\protect\cite{Ashfold:1985,Ziegler:1987}.  However, the much stronger
rotational dependence observed in the RRREP's could have also been described by
Coriolis forces\protect\cite{Ziegler:1987}.  These two different mechanisms
could
not be distinguished since the results of these experiments represent an
average increase among all K sublevels such that distinguishing between the two
different mechanisms is difficult.

Fully rotationally resolved spectra have been recorded by optical-optical
double resonance methods\protect\cite{Ashfold:1986a,Xie:1986}.
However, these experiments were not able to establish a significant
predissociation dependence on rotational level.  It should be noted that one
step of this double resonance scheme involves a multiphoton excitation that
might be expected to distort the observed lineshapes.  Thus, due to the
experimental uncertainty of the determined linewidths in these experiments,
a weak predissociation dependence on rotational level might have been
masked.

Microwave detected, microwave optical double resonance has recently
been used to probe the dynamics in the NH$_{3}$ \~A state 2$^{0}$
 band\protect\cite{Endo:1990}.   The rotational dependence observed in this
level was again
quantitatively fit using the centrifugal model\protect\cite{Ashfold:1991}.
  In the previous paper, we
reported the \~A $\leftarrow$ \~X transition frequencies measured by MODR for
the higher
lying vibrational levels of the NH$_{3}$ \~A state as well as for the
isotopomers\protect\cite{Henck:1994a}.
This technique has also enabled us to measure the
homogeneous linewidths for individually resolved $J_{K}$ rovibronic transitions
for the entire isotopomeric series.  Our observed linewidths are considerably
narrower than previous reports\protect\cite{Ashfold:1986a,Xie:1986}.
 Thus, it was hoped that these results would
enable us to discern for the various isotopomers the vibrational and rotational
contributions to the \~A state predissociation mechanisms.

Assuming that stretching motions are more correctly described by a
local mode coordinate system rather than a normal mode one, a crude one
dimensional potential was designed along an N-H(D) internal bond
coordinate.  This potential has been used to demonstrate that a 2-1 Fermi
resonance between $\nu_{2}$ and $\nu_{1}$ quantitatively predicts the observed
linewidths
for levels with more than one quantum in $\nu_{2}$.  For the lowest two levels,
predissociation could only occur by tunneling through this barrier.  To
reproduce the observed linewidths of the vibrationless levels in NH$_{3}$ and
ND$_{3}$, the barrier height was reduced roughly 1000~cm$^{-1}$ relative to the
{\it ab initio\/} value.  This result agrees with the conclusion of
Dixon\protect\cite{Dixon:1989}
 who also has demonstrated that the {\it ab initio\/}
calculations\protect\cite{McCarty:1987}
overestimate the barrier.  The predissociation efficiency was also shown
to be weakly dependent upon rotation.  Surprisingly, this rotational
dependence was found to be isotope dependent.  For the hydrogen containing
isotopomers, the rotational dependence was more adequately described by
Coriolis forces coupling energy from $\nu_{2}$ and into $\nu_{3}$
and $\nu_{4}$, while for ND$_{3}$,
this rotational dependence obeyed the centrifugal model.  This difference
arises from the fact that in NH$_{3}$, one quantum of $\nu_{2}$ added to the
N--H zero
point energy is isoenergetic with scattering states above the barrier to
dissociation while for ND$_{3}$, this energy still lies below the top of the
barrier.  Thus, for ND$_{3}$, the Coriolis contribution to the predissociation
rate is
reduced such that tunneling remains the dominant predissociation pathway.

\section{Experimental}

	The recording and analysis of the MODR and RFODR data was described in
the previous paper and will not be repeated here.  The linewidths obtained
from such an analysis are given in tables
\protect\ref{NH3Widths}, \protect\ref{NH2DWidths}, \protect\ref{NHD2Widths},
 and \ref{ND3Widths}.
As suggested by Ziegler\protect\cite{Ziegler:1987}, the
observed widths were fit to,
\begin{equation}
\Gamma = \Gamma_{o}\times exp\{\Gamma_{B}[J(J+1)-K^{2}] +\Gamma_{C}K^{2}\}
\label{Fit_Equation}
\end{equation}
where $\Gamma$ is the observed width, $\Gamma_{o}$ is a determined constant for
a given
vibronic band, $\Gamma_{B}$ reflects the enhancement by rotation in the plane
of the molecule, and $\Gamma_{C}$ that of the enhancement due to rotation
about the symmetry axis.  For the asymmetric top species, K$^{2}$ is replaced
by
$\langle J^{2}_{z}\rangle /K^{2}$, calculated from the fitted rotational
constants.  In these fits, the observed widths were weighted by
the reciprocal of the squared uncertainty predicted by the lineshape fit.
Results are given in table~\ref{DataFit}. Because the
microwave probe transitions are Q branch lines of the inversion band,
our data is primarily for high K lines.  For
the mixed isotopomer data, we were not able to determine the rotational
dependence of the widths due to the limited number of transitions observed
in these molecules.   For the asymmetric tops, the ground state Q branch is
widely spread, reducing the number of strong transitions in the region
covered by our microwave equipment.

\section{Discussion}

\subsection{Empirical Local Mode Potential Surface}

	The \~A state absorption spectrum has been modeled as a
two symmetric mode problem\protect\cite{Rosmus:1987,Tang:1988}.
One quantum of stretching excitation
contains more energy than needed to overcome the barrier along a single
 N--H bond, yet the $\nu_{1}$ normal mode coordinate is still strongly bound.
The
dissociation reaction coordinate is a mixture of $\nu_{1}$ and the two
components of
the antisymmetric N--H stretching mode, $\nu_{3}$.   This potential topology
will lead
to strong nonlinear interaction between the $\nu_{1}$ and $\nu_{3}$ normal
modes such
that an initial state corresponding to $\nu_{1}$ excitation will rapidly
trifurcate into
three equivalent wave packets corresponding to bond-breaking along each of
the three equivalent N--H(D) bonds.  The depth of the exit channels and the
sharpness of the motion along the symmetric stretch should lead to little
amplitude being reflected back. Thus, one should expect only weak
recurrence\protect\cite{Tannor}.

	In work on stable symmetric hydrides, such as the ground state of
ammonia, it has been established that one can treat the X--H stretching motion
equivalently with either a local mode or normal mode basis.
 When the single bond anharmonicity dominates over the bond-bond harmonic
coupling, the eigenstates are expected to be close to symmetrized local mode
states.  Contrarily, when the harmonic coupling dominates over
anharmonicity, the eigenstates are closer to normal mode basis functions.  For
perfect harmonic motion, the local mode and normal mode descriptions of
the fundamental levels are identical.  For the ammonia \~A
state, however, the potential along a single N--H bond is highly anharmonic.
In fact, as will be shown below, the zero point level is the only resonance
below the barrier.  As a result, excitation of one bond mode leads to
dissociation in a single vibrational period \mbox{(10 fsec)}.  The time
required for
exchange of vibrational energy between the N--H bonds is longer by
approximately $f_{rr}$/$f_{rr'}$, which based upon the {\it ab initio} harmonic
force field%
\protect\cite{Rosmus:1987}  is predicted to be greater than 50.   Thus, the
\~A state of ammonia is the `ultimate' local mode molecule,
with bond-bond coupling negligible compared to the rate of dissociation if the
N--H stretching modes are excited.  As a result, we have modeled the
dissociation in terms of three uncoupled, single bond modes.

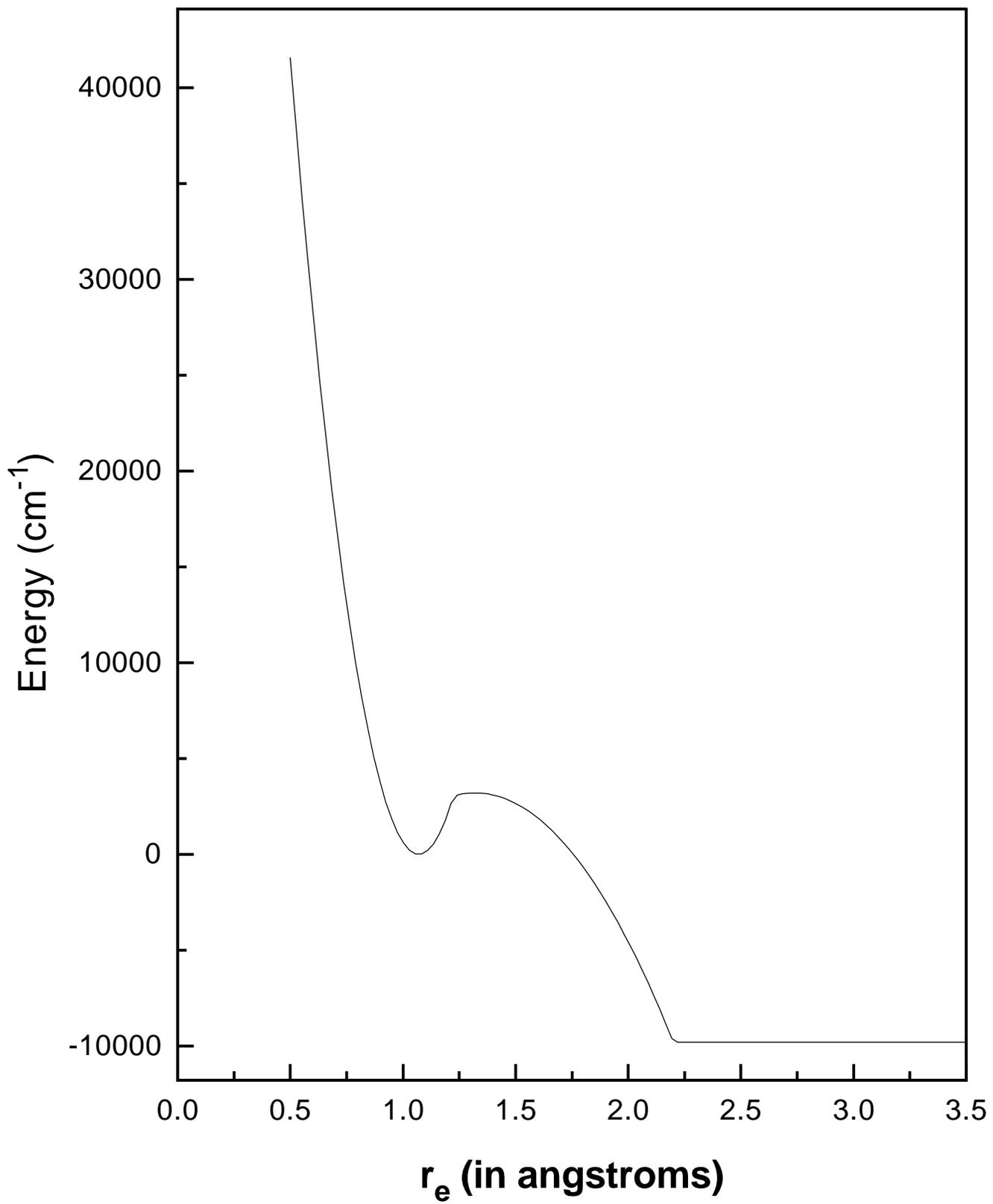
\begin{figure}
\caption{Model one-dimensional energy surface of the \~A state of
ammonia along the dissociative N--H coordinate.  Curve is composed
of two parabolas with positions and radius of curvature
corresponding to the calculated values.}
\label{Potential}
\end{figure}

	Figure~\ref{Potential} shows a model potential we have used to fit the
observed
dissociation rate.  The curve contains information from the {\it ab initio\/}
calculations\protect\cite{McCarty:1987} as well as from the experimentally
measured information
about the position of the barrier\protect\cite{Henck:1994a}. These values are
repeated below.
\vskip.1in
\begin{center}
\begin{tabular}{|c|r|}\hline
$r_{e}$ & 1.06 \AA\\
$r_{b}$ & 1.32 \AA\\
$\Delta H$ & 3200 cm$^{-1}$ \\
$f_{rr}(r<r_{b})$ & $2.56\times 10^{5} cm^{-1}/\AA^{2}$ \\
$f_{rr}(r_{b}<r<R)$ & $-3.36\times 10^{4} cm^{-1}/\AA^{2}$ \\ \hline
\end{tabular}
\end{center}
\vskip0.1in
\noindent The curve consists of a pair of parabolas whose extrema and curvature
are based
upon the {\it ab initio\/} potential\protect\cite{McCarty:1987}, but whose
barrier
height can be adjusted.  Beyond some large separation, $R \gg  r_{e}$,
the potential was taken to
be -10000~cm$^{-1}$ correlating with the energy difference of the NH$_{3}$
\~A state and
the ground states of the photofragments, NH$_{2}$ and
H\protect\cite{Runau:1977}.
The real eigenstates, $\psi$(E) were calculated at each energy by numerical
integration and normalized such that
$\int \psi (E)\psi(E^{'})dx = 2\pi\delta(E-E^{'})$.

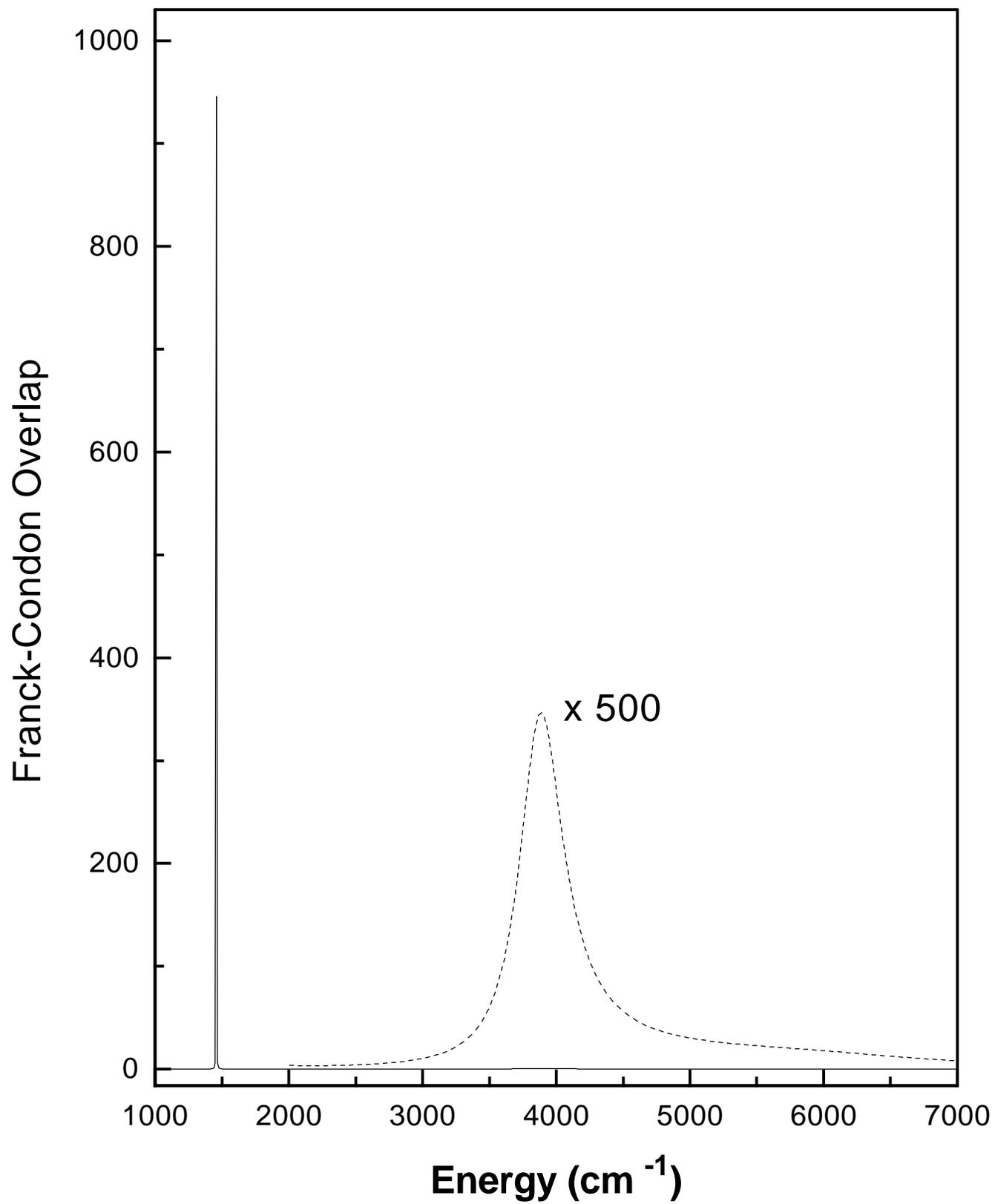
\begin{figure}
\caption{Predicted Franck-Condon overlap for N--H from ground state
wavefunction.  Upper
state potential has a 3200~cm$^{-1}$ high barrier to dissociation.  First
excited
state lies above the barrier and lies 2480~cm$^{-1}$ above the zero point level
and has $<$1\% of the peak height.}
\label{Ab_Initio_Overlap}
\end{figure}

	We begin our discussion with the barrier height picked to agree with
the most recent {\it ab initio\/} prediction of
3200~cm$^{-1}$\protect\cite{McCarty:1987}.
In order to mimic what we would expect
for $\nu_{1}$ absorption activity, we
calculated the Franck-Condon overlap of $\psi$(E) with a Gaussian picked to
model the zero point level of the ground electronic state.  The resulting
squared overlap versus energy is presented in figure~\ref{Ab_Initio_Overlap}.
We observe a sharp
resonance at 1450~cm$^{-1}$ with a HWHM of 0.16~cm$^{-1}$.  This resonance
corresponds to the zero-point level below the dissociation barrier.  If this
represented the vibrationless level of the \~A state, one predicts a
homogeneous width of three times this size since there are three equivalent
hydrogen atoms that could tunnel out; thus, the lifetime broadening will be
three times the lifetime broadening of each N-H bond.  This prediction is
much smaller than the 17~cm$^{-1}$ HWHM measured for the NH$_{3}$ origin.

	While the features of our crude potential have a clear physical
interpretation, this potential overestimates the thickness of the barrier
relative to the {\it ab initio\/} potential\protect\cite{McCarty:1987}.  As a
consistency check, using a spline
interpolation, we numerically solved the one-dimensional Schroedinger
equation for this potential.  Franck-Condon overlap with the ground state
wavefunction resulted in a sharp resonance at 1460~cm$^{-1}$ with a width of
0.73~cm$^{-1}$ in qualitative agreement with the results from our more crude
approach.  Thus, we justify our simple method and conclude that the {\it ab
initio\/}
calculation has overestimated the barrier height.   Dixon reached the same
conclusion based upon his two-dimensional wavepacket
calculations\protect\cite{Dixon:1988}.

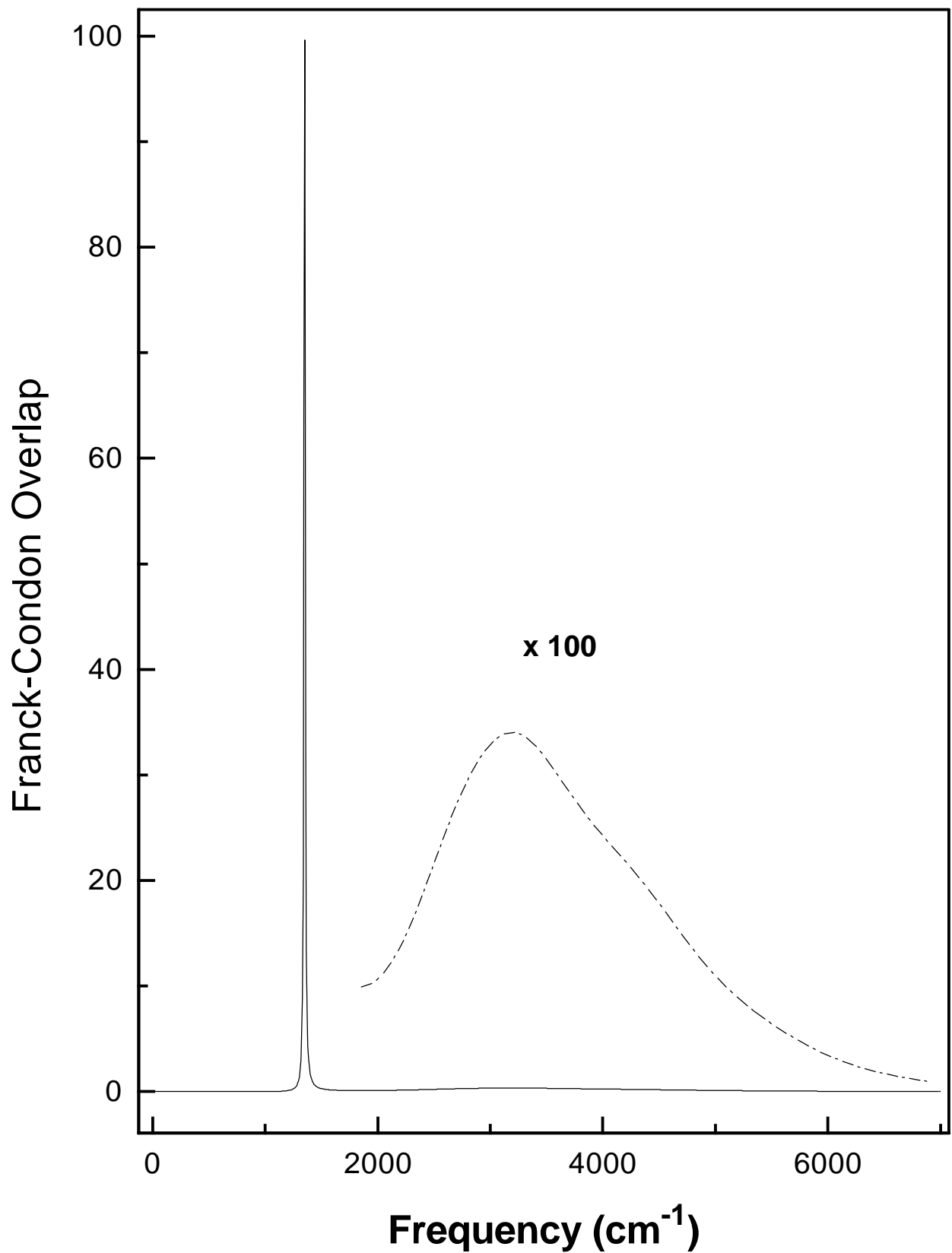
\begin{figure}
\caption{Predicted Franck-Condon overlap for N--H from ground state
wavefunction as a
function of energy above $T_{e}$.  Upper state potential has a 2075~cm$^{-1}$
high barrier
to dissociation.  The first excited state lies appoximately 2000~cm$^{-1}$
above the zero point
level and is approximately 950~cm$^{-1}$ wide.}
\label{NH3_Overlap}
\end{figure}

	In order to reproduce the observed  17.1~cm$^{-1}$ width of the NH$_{3}$ zero
point transition, the barrier height in our model potential was reduced to
2075~cm$^{-1}$.  Plotted in figure~\ref{NH3_Overlap} is the Franck-Condon
overlap as a function of
energy above $T_{e}$ for this corrected potential.  The zero point level lies
1350~cm$^{-1}$
above $T_{e}$ and has a HWHM of 5.7~cm$^{-1}$ per hydrogen in good agreement
with the experimental measurement. Thus, in order to model the
experimentally measured width of the NH$_{3}$ zero point level, we need to
lower the {\it ab initio\/} barrier height\protect\cite{McCarty:1987}
 by more than 1000~cm$^{-1}$.

	As shown in figure~\ref{NH3_Overlap}, an extremely  broad quasi-resonance
above the
barrier is also present.  This quasi-resonance represents what may be called
one quantum in the N-H local mode.  Eigenstates in this energy region have
their inner turning points at bond lengths slightly less than $r_{e}$, and thus
have significant overlap with the ground state Gaussian.   Our crude
calculation predicts this resonance lies 3370~cm$^{-1}$ above $T_{e}$
(2020~cm$^{-1}$ above
the zero point level) and has a 2850~cm$^{-1}$ linewidth.  Because of its broad
width, the peak absorption is expected to be small relative to the zero point
level, making this level very difficult to detect.  Ashfold {\it et
al.}\protect\cite{Ashfold:1986a}
have measured a 500~cm$^{-1}$ wide resonance centered at 2300~cm$^{-1}$
above the zero point level that they have assigned to 1$^{1}$ .
As discussed in our previous paper, we have reassigned this band to 4$^{2}$.
This level is very likely to be strongly
coupled to the N--H stretching modes through Fermi coupling leading to the
observed rapid dissociation rate.

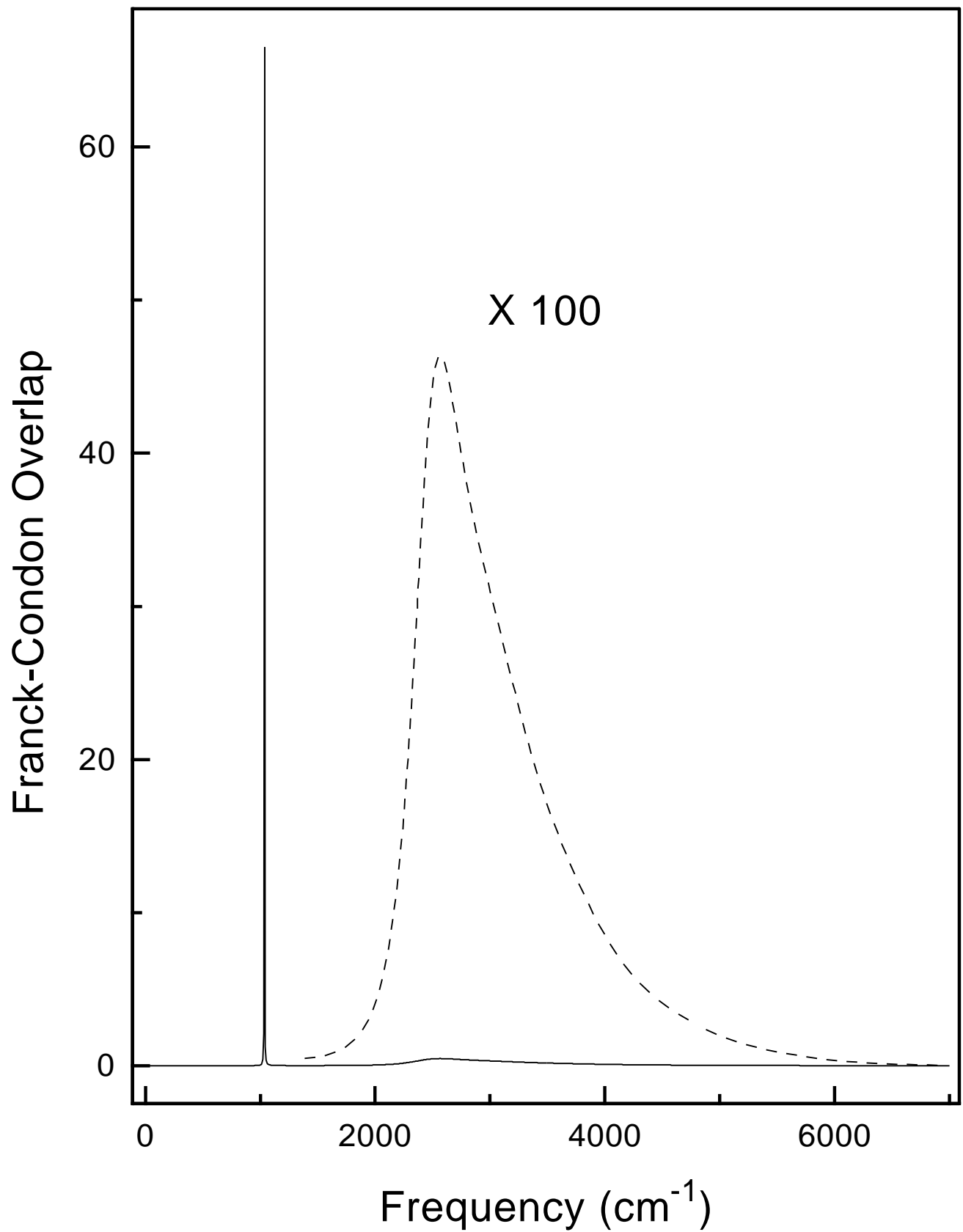
\begin{figure}
\caption{Predicted Franck-Condon overlap for N--D from ground state
wavefunction as a
function of energy above $T_{e}$.  Upper state potential has a 2075~cm$^{-1}$
high barrier
to dissociation.  The furst excited state lies appoximately 1700~cm$^{-1}$
above the zero point
level and is approximately 450~cm$^{-1}$ wide.}
\label{ND3_Overlap}
\end{figure}

	Using the same potential, we also calculated the Franck-Condon
overlap for an N--D bond.  Results from the calculation are presented in
figure~\ref{ND3_Overlap}.
The zero point level is found at 1035~cm$^{-1}$ with a width of 0.16~cm$^{-1}$
per
deuterium.  The ND$_{3}$ zero point level is predicted to have a homogeneous
linewidth of 0.48~cm$^{-1}$ comparing favorably with the experimental
measurement of 0.4~cm$^{-1}$. We calculate the N-D second resonance to have
a homogeneous linewidth of 1360~cm$^{-1}$ and to lie about
2750~cm$^{-1}$ above $T_{e}$ or
about 1720~cm$^{-1}$ above the zero point level.  A broad feature was observed
at
1790~cm$^{-1}$ in the ND$_{3}$ dispersed fluorescence spectrum that has been
assigned
to the 1$^{1}$ band.  As for NH$_{3}$, we reassign this feature to the 4$^{2}$
band.

	A quantitative calculation to test the accuracy of this local mode
approximation is clearly desirable.  Unfortunately, this calculation would
require at a minimum {\it ab initio\/} calculations along the three stretching
coordinates, and then a 3-D wavepacket propagation of the ground state
Gaussian on this potential.  If such a calculation dramatically reduces the
predicted resonance widths of the 1$^{1}$ level, then we would withdraw our
reassignment of the  peaks Ashfold {\it et al.}\protect\cite{Ashfold:1986a}
assign
to 1$^{1}$.  However, since the basic vibronic and isotopic trends are
predicted by
this crude potential, we believe our empirical dissociation potential contains
the essential features of the true potential.

\subsection{Predissociation dependence upon vibrational level}

	The increasing dissociation rate with $\nu_{2}$ excitation above 2$^{1}$
suggests
that anharmonic interactions are coupling two quanta of vibrational energy
out of the bending normal mode into one of the three N-H(D) bond
coordinates lying along the dissociation coordinate\protect\cite{Vaida:1987}.
The
lowest order vibrational coupling that could be responsible is  a Fermi
resonance between $\nu_{2}$ and $\nu_{1}$ caused by the
$k_{hhs}$ term in an harmonic
approximation.  We estimate the value of this coupling constant by assuming
the molecule vibrates out-of-plane with constant bond length.  This
approximation is justified by the {\it ab initio\/}
potential\protect\cite{McCarty:1987}.

	Classically, the above coupling leads to a forcing of the stretching mode
at two times the frequency of the bending mode.  If there is a 2:1 resonance of
the harmonic frequencies, one would expect an efficient transfer of energy
inside some resonance zone.  The experimentally determined force field
given in the previous paper does not predict an exact
resonance\protect\cite{Henck:1994a},
but a small amount of average energy transfer would be expected due to
non-resonant
coupling.  Further, given the rapid dissociation of the N-H(D)
stretching modes, their frequencies are poorly defined, making the resonance
condition much less restrictive.  A 2:1 Fermi resonance implies the transfer of
two quanta in $\nu_{2}$ into one quantum in $\nu_{1}$; anharmonic transfer of
one
quantum of $\nu_{2}$ is rigorously symmetry forbidden.  We see that this
mechanism operates only for levels containing two or more quanta in $\nu_{2}$
and
the rapid increase in linewidth observed above the 2$^{1}$ level is predicted.

	We use our empirical potential to estimate the expected magnitude of
the Fermi resonance.  We use as a zero order model, the coupling of two
harmonic oscillators and by second order perturbation theory, we find that
the average vibrational energy in the stretching coordinate increases by an
amount given by,
\begin{equation}
\langle \Delta
E_{s}\rangle=\omega_{s}\biggl(\frac{k_{hhs}^{2}}{4\sqrt{2}}\biggr)
\biggl[\frac{\nu_{2}(\nu_{2}-1)}{(\omega_{s}-2\omega_{2})^{2}}+
\frac{(2\nu_{2}+1)^{2}}{\omega^{2}_{s}}+
\frac{(\nu_{2}+1)(\nu_{2}+2)}{(\omega_{s}+2\omega_{2})^{2}}\biggr]
\label{Perturbation_Formula}
\end{equation}
where $\omega_{s}$ is the stretching frequency, $\omega_{2}$ is the bending
frequency,
$\nu_{2}$ is the number of quanta in the out-of-plane bend, and $k_{hhs}$
is the coupling constant.  Using the results from
Ziegler\protect\cite{Ziegler:1987},
the effect of the Fermi resonance on the transition linewidth is
\begin{equation}
\Delta \Gamma = \Gamma_{o} \times exp\biggl[
\frac{\langle \Delta E_{s}\rangle}{\hbar \omega_{2}}\biggr]
\label{Ziegler}
\end{equation}
where $\hbar\omega_{2} = 1063$~cm$^{-1}$ is the imaginary frequency at the top
of the barrier.
Putting in values estimated from the {\it ab initio\/} force
field\protect\cite{Rosmus:1987}
gives an increase in energy that is far too small to explain the observed
vibrational dependence of the widths.

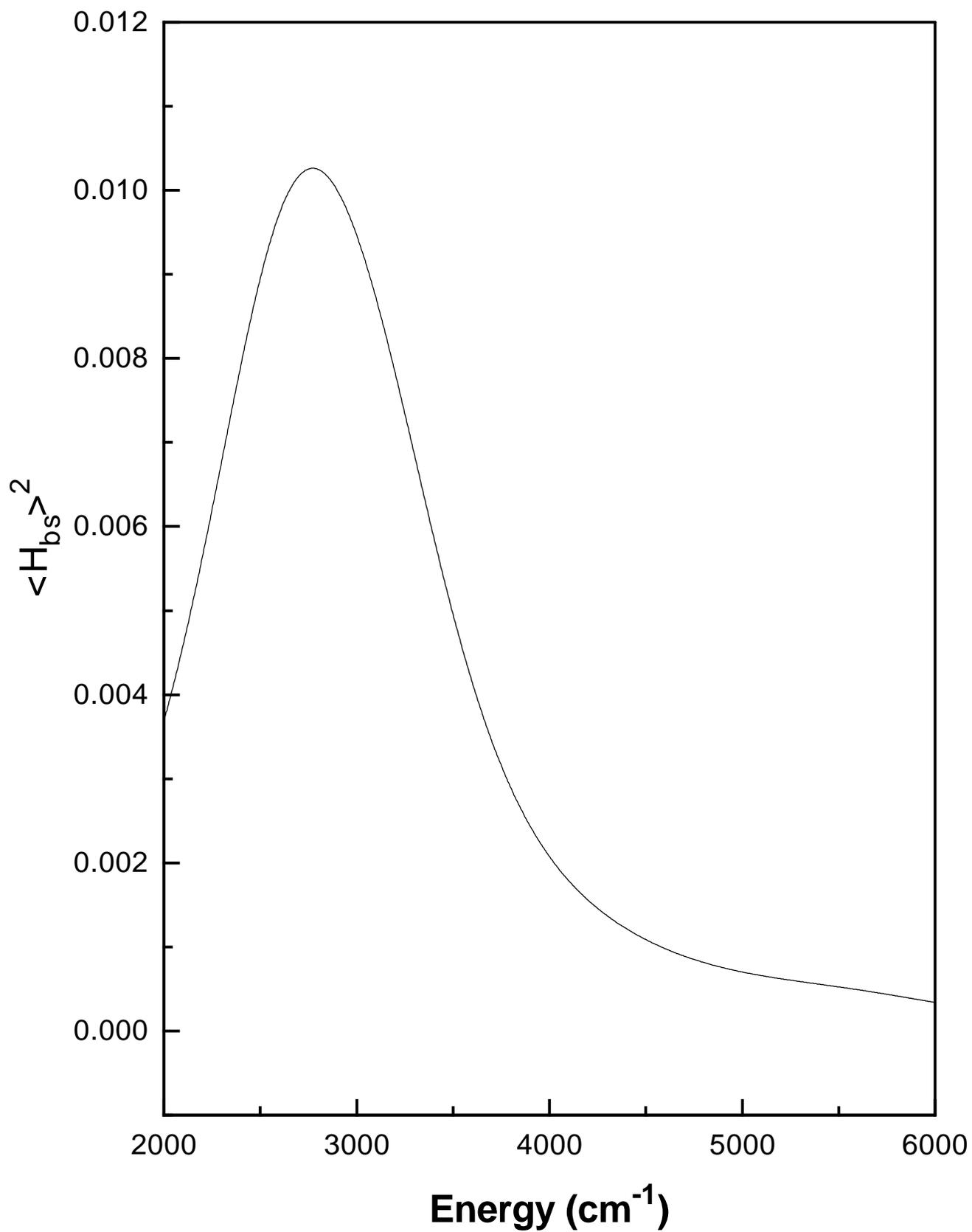
\begin{figure}
\caption{The squared bend-stretch coupling matrix element for the bend and the
stretch
versus the
energy in a single N--H stretch including the zero point energy.}
\label{HFermi}
\end{figure}

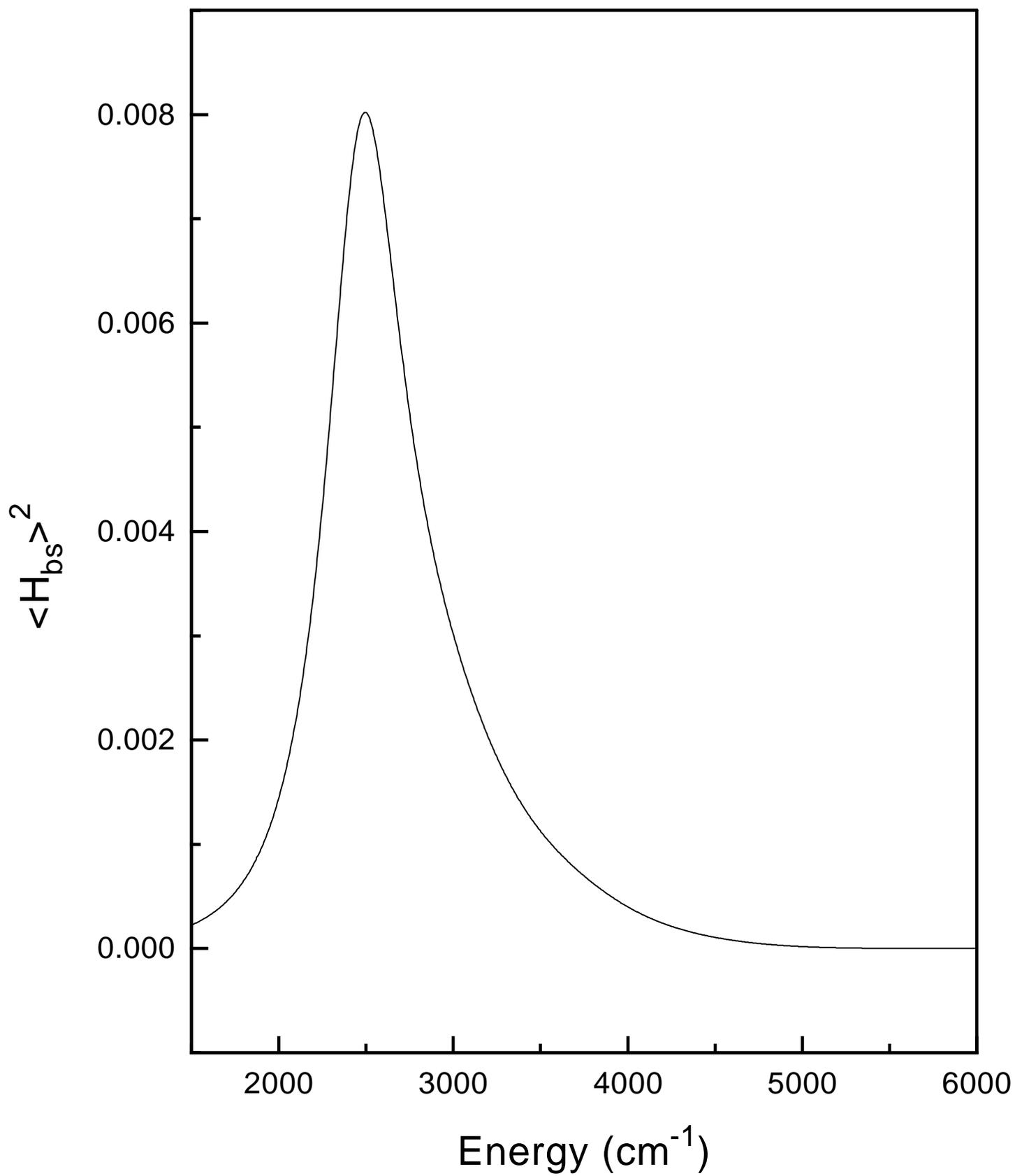
\begin{figure}
\caption{The squared bend-stretch coupling matrix element for the bend and the
stretch
versus the
energy in a single N--D stretch including the zero point energy.}
\label{DFermi}
\end{figure}

	In order to account for the highly anharmonic motion along the N--H(D)
stretching coordinates, we use our above estimate for this potential
curve.  States with two quanta in the out-of-plane bending mode couple
directly to scattering states above the dissociation barrier.  We use Fermi's
Golden rule to estimate the contribution of the coupling to the dissociation
rate,
\begin{equation}
k_{diss}=\frac{4\pi}{\hbar^{2}}\sqrt{\frac{2m}{E}}\langle V_{C} \rangle^{2}
\label{Golden_Rule}
\end{equation}
where $\langle V_{C}\rangle$ is the coupling matrix, $E$ is the energy relative
to the bottom of the
exit channel, and the $\sqrt{2m/E}$ factor arises due to the density of
states and box normalization of the wavefunctions.  Assuming the simplest
resonance,
Fermi coupling between the bend and the stretch is predicted by
\begin{equation}
\langle V_{C}\rangle=\frac{1}{2}f_{rhh}\langle
\nu_{2}-2|h^{2}|\nu_{2}\rangle\langle
E_{o}+\Delta E_{2}|(r-r_{o})|\Psi_{o}\rangle^{2}
\label{Fermi_Coupling}
\end{equation}
where $f_{rhh}$ is the cubic bend stretch force constant, h is the out-of-plane
bend
internal coordinate, and $\langle E_{o}+\Delta E_{2}|(r-r_{o})|\Psi_{o}\rangle$
is the coupling matrix element. $|\Psi_{o}\rangle$ is the zero point resonance
normalized to unity inside the well and $\langle E_{o}+\Delta E_{2}|$ is a
continuum function normalized to $2\pi sin(kr)$ in
the asymptotic region ($r \gg r_{o}$).  We use the harmonic approximation for
the
out-of-plane bend and values for $f_{hh}$ from our experimentally estimated
force field to determine the expectation value for $h^{2}$.  This value was
given by
\begin{equation}
\langle \nu_{2}-2|h^{2}|\nu_{2}\rangle =7.614\times
10^{-3}\sqrt{\nu_{2}(\nu_{2}-1)}\AA^{2}
\label{H2_Matrix_Element}
\end{equation}
Given in figures~\ref{HFermi} and ~\ref{DFermi} is the squared
coupling matrix element as a function of the energy in the N--H(D) bond.
Collecting the above expressions along
with the {\it ab initio\/} value for $f_{rhh}$ of 2.66
aJ/\AA$^{3}$\protect\cite{Rosmus:1987}
allowed estimation of the
predissociation rates and thus the expected transition linewidths.  These
calculated values are presented in table~\ref{Linewidths} along with our
measured values.

	The predicted dissociation rates are about a factor of three too large for
the 2$^{2}$ levels, and increase too rapidly above that.  Roughly, the isotopic
dependence is correctly predicted.  One explanation for the predicted
dissociation being much faster than observed is that the {\it ab initio\/}
value of $f_{rhh}$
may be too large.  Even more significant may be our neglect of the
barrier dependence on out-of-plane angle.  Increasing the
barrier height will reduce the coupling matrix element among the zero point
level and the scattering states and decrease the predicted Fermi resonance
contribution to the dissociation rate.  Consequently, we conclude that
our overestimate of the Fermi contribution to the dissociation rate is caused
by the neglect of the barrier dependence on out-of-plane deformation.
Proper treatment of this
problem, including both Fermi resonance and the effect of barrier changes
with out-of-plane angle, requires two-dimensional wavepacket calculations
that are beyond our computational resources.

	This resonance calculation does rationalize the larger than expected
dissociation rate for the NHD$_{2}$ isotopomer.  Even though there is great
uncertainty as to the frequency of the stretching levels, two quanta of
$\nu_{2}$ for
NHD$_{2}$ would be expected to be in closer resonance with the N--D stretch
than
for two quanta in any of the other isotopomers, leading to an enhanced
dissociation rate. Contrary to the mechanism controlling the dynamics in
the tunneling region, this mechanism predicts that the N--D bond dissociates
faster than the N-H bonds for these higher lying levels.

\subsection{Predissociation dependence on rotational level}

	The predissociation dependence on rotational level can arise from two
independent sources.  In the first mechanism, as extensively discussed by
Ashfold et al.\protect\cite{Ashfold:1986a,Ashfold:1985}  and
Ziegler\protect\cite{Ziegler:1987},
rotation about either the b or c axes will
promote an N--H(D) bond lengthening resulting in an effective barrier
reduction.
In the second mechanism, b axis rotation causes the $\nu_{2}$ mode to be
coupled to
the unbound $\nu_{3}$ and $\nu_{4}$ modes through Coriolis forces and, hence,
increasing
the probability of barrier penetration.  If centrifugal effects are the
dominant
source of rotational enhancement, the ratio of b to c axis effects should be
constant with vibronic excitation.  Contrarily, the Coriolis
coupling matrix elements would scale with $\nu_{2}$.  Hence, the two
mechanisms can be distinguished by measuring the ratio of b to c axis effects
as a function of $\nu_{2}$.

	We estimate this ratio allowing only centrifugal effects to contribute.
Assuming the dissociating H(D) atom lies on the y axis, the effective potential
is given by
\begin{equation}
V({\roarrow{r}})=V_{o}({\roarrow{r}})+
\frac{[J(J+1)-K^{2}]}{4I_{x}({\roarrow{r}})}
+\frac{K^{2}}{2I_{z}({\roarrow{r}})}
\label{Centrifugal}
\end{equation}
where $V_{o}(\roarrow{r})$ is the rotationless potential and $1/2[J(J + 1) -
K^{2}]$
and $K^{2}$ are the
expectation values for x and z axis rotation respectively.  Using the {\it ab
initio\/}
values for $r_{e}$ and $r_{b}$ at 1.055~\AA\ and 1.32~\AA\ respectively,
the rotational correction to the barrier height is estimated to be
\begin{equation}
\Delta V=1.25[J(J+1)-K^{2}]+0.73K^{2}.
\label{Rotational_Correction}
\end{equation}
Thus, the centrifugal model predicts a B axis effect to C axis effect ratio of
1.75.
We also note that for the centrifugal model this ratio is independent of the
barrier shape and only weakly dependent upon barrier position.  For instance,
as $r_{b}$ was varied from $r_{e}$ to $\infty$, this ratio only changed from 2
to 1.

	From our NH$_{3}$ data, the ratio of b axis to c axis effects was found to be
2.2 $\pm$ 1. for the zero point
level and 15. $\pm$ 2. for the 2$^{1}$ level.  We predict that centrifugal
effects dominate
in the vibrationless level while Coriolis forces dominate in the 2$^{1}$ level.
 This
change in dominant mechanism can be explained utilizing our empirical N--H
local mode potential.  This potential predicts the zero point energy of an N--H
bond mode to be 1350~cm$^{-1}$ while the barrier height was determined to be
2075~cm$^{-1}$.  The 2$^{1}$ level lying roughly 900~cm$^{-1}$
above this level is expected to
lie well above the barrier and therefore be Coriolis coupled to a manifold of
continuum states.

	Using a procedure similar to the one we used to estimate the
magnitude of the Fermi resonance, we estimate the rotational dependence of
the linewidths in the NH$_{3}$ 2$^{1}$ level due to Coriolis forces.  In this
case, the
coupling matrix elements are given by
\begin{equation}
\langle V_{C}\rangle = -I_{xx}^{-1}\frac{\hbar}{2}
\sqrt{J(J+1)-K^{2}}\zeta_{sb}^{x}\langle Q_{s}P_{b} - Q_{b}P_{s}\rangle
\label{Coriolis_Matrix_Element}
\end{equation}
where $I_{xx}$ is the moment of inertia about the x axis, $\zeta_{sb}^{x}$
is the stretch-bend Coriolis coupling constant which we set to
its ground state value of 0.752, $Q_{s}$ and $Q_{b}$ are the
normal mode coordinates for the stretch and bend, and $P_{s}$ and $P_{b}$
are the conjugate momenta to these
normal coordinates.  By substituting equation~\ref{Coriolis_Matrix_Element}
into equation~\ref{Golden_Rule}, we can
estimate the Coriolis enhancement to the predissociation rate and the
expected linewidth enhancement.  The Coriolis contribution to the linewidth
was calculated to be
\begin{equation}
\Gamma_{o} = 0.11 [ J (J + 1) - K^{2} ]
\end{equation}
For the NH$_{3}$ 2$^{1}$ level, this predicted value is in excellent agreement
with the
observed b axis rotational dependence of 0.10(4)~cm$^{-1}$.

	The ND$_{3}$ results are more poorly determined than our NH$_{3}$ results.
We still can compare our determined ratio of b axis to c axis enhancement for
the two isotopomers.  In contrast to NH$_{3}$, this ratio for the
ND$_{3}$ 2$^{1}$ level was
determined to be 2.1(7) more in line with the centrifugal model.  Surprisingly,
for the two isotopomers, the dominant mechanism for rotational
enhancement appears to be different.

	To understand this result we again turn to our empirical N-H local
mode potential.  This potential predicts the zero point energy in the N--D
stretch to be 1035~cm$^{-1}$.  Addition of one quantum in $\nu_{2}$ with an
energy of
approximately 670~cm$^{-1}$ still lies well below the barrier.  Since continuum
states below the top of the barrier have small amplitude near $r_{e}$, the
Coriolis
coupling matrix element should be much smaller for ND$_{3}$ than for NH$_{3}$.
 Therefore, changes in the zero
point energies and mode frequencies due to isotopic substitution are shown
to have a dramatic effect on the predissociation dependence on rotational
level.

\section{Conclusion}

	We have measured the homogeneous linewidths of individual
rovibrational transitions using MODR.  Our measured linewidths are
considerably narrower than previous
determinations\protect\cite{Ashfold:1986a,Xie:1986}.
Since the observed widths were much larger than the laser bandwidth or the
Doppler width, they
represent a direct measure of the predissociation lifetime.  Thus, we have
used these linewidths to more fully interpret the \~A state
predissociation mechanisms.

	Using the available experimental and theoretical data, we have
developed a one-dimensional, local mode potential along a N--H(D) bond.
This potential has been used to qualitatively predict the vibronic and isotopic
dependence of the predissociation rates.  In order to simulate the observed
widths in the zero point levels, the barrier height had to be lowered to
2075~cm$^{-1}$,
about 1000~cm$^{-1}$ below the {\it ab initio\/}
prediction\protect\cite{McCarty:1987}.
Dissociation rates from
the 2$^{0}$ and 2$^{1}$ levels could only proceed by tunneling through this
barrier
while dissociation from the higher lying levels was described by a 2:1 Fermi
resonance of $\nu_{2}$ directly to the manifold of N--H(D) continuum states
lying above
the barrier.   This model also predicts that photdissociation of NHD$_{2}$
should
proceed primarily through the $NHD + D$ pathway.

	Dissociation of the two lowest vibronic bands was found to have a
weak dependence upon rotation level.  Rotationally enhanced dissociation
from the zero point level in both NH$_{3}$ and ND$_{3}$ is caused by the
lowering of
the effective barrier height by centrifugal forces
.  The rotational enhancement of the
dissociation rate from the NH$_{3}$ 2$^{1}$ level, which is predicted to lie
above the
potential barrier along the N--H stretch, was dominated by perpendicular
Coriolis coupling to the manifold of continuum states lying above the barrier.
The same level in ND$_{3}$ lies below the barrier to dissociation and the
rotational dependence of the dissociation rates still fits the centrifugal
model\protect\cite{Ashfold:1985}.

\begin{table}
\caption{Linewidths of the NH$_{3}$ \~A $\leftarrow$ \~X state MODR
transitions.}
\begin{center}
2$^{0}$ band.
\begin{tabular}{cdd}
{\bf Transition}  &  {\bf HWHM (cm$^{-1}$)}  &  {\bf {1$\sigma$}}  \\ \hline
Q$_{2}$(2)  &   16.83&0.06\\
R$_{2}$(2)  &   19.38&0.21\\
Q$_{3}$(3)  &   17.90&0.05\\
R$_{3}$(3)  &   19.37&0.22\\
Q$_{4}$(4)  &   18.81&0.11\\
Q$_{4}$(5)  &   18.11&0.25\\
R$_{4}$(5)  &   24.20&3.60\\
Q$_{5}$(5)  &   19.19&0.13\\
R$_{5}$(5)  &   25.98&0.99\\
Q$_{6}$(6)  &   19.48&0.08\\
R$_{6}$(6)  &   33.31&1.52\\
P$_{6}$(7)  &   16.27&1.57\\
Q$_{6}$(7)  &   20.79&0.14\\
R$_{6}$(7)  &   26.48&1.37\\
Q$_{7}$(7)  &   21.04&0.14\\
R$_{7}$(7)  &   23.70&1.88\\
Q$_{8}$(8)  &   21.00&0.11\\
R$_{8}$(8)  &   26.03&2.35\\
Q$_{9}$(9)  &   22.74&0.44
\end{tabular}
2$^{1}$ band.
\begin{tabular}{cdd}
{\bf Transition}  &  {\bf HWHM (cm$^{-1}$)}  &  {\bf {1$\sigma$}}  \\ \hline
Q$_{1}$(1)    & 15.03&0.07\\
R$_{1}$(1)    & 14.82&0.15\\
P$_{2}$(3)    & 12.73&0.16\\
Q$_{2}$(3)    & 15.39&0.05\\
R$_{2}$(3)    & 15.99&0.08\\
Q$_{3}$(3)    & 15.07&0.02\\
R$_{3}$(3)    & 15.48&0.09\\
P$_{2}$(5)    & 13.88&0.33\\
Q$_{2}$(5)    & 15.56&0.45\\
R$_{2}$(5)    & 17.24&0.27\\
P$_{4}$(5)    & 13.64&0.14\\
Q$_{4}$(5)    & 19.65&0.16\\
R$_{4}$(5)    & 16.35&0.06\\
Q$_{4}$(5)    & 14.75&0.43\\
R$_{4}$(5)    & 17.68&0.31\\
P$_{3}$(6)    & 18.27&0.35\\
Q$_{3}$(6)    & 12.10&0.81\\
R$_{3}$(6)    & 16.75&0.08 \\
P$_{5}$(6)    & 18.98&0.54\\
Q$_{5}$(6)    & 11.09&0.54\\
R$_{5}$(6)    & 17.14&0.08\\
R$_{6}$(7)    & 21.28&0.54\\
Q$_{7}$(8)    & 17.93&0.15\\
Q$_{8}$(8)    & 16.04&0.03\\
R$_{8}$(8)    & 20.10&0.79\\
P$_{8}$(9)    & 46.17&6.61\\
R$_{8}$(9)    & 23.22&2.33\\
Q$_{9}$(9)    & 14.64&0.09\\
R$_{9}$(9)    & 33.74&5.11
\end{tabular}
2$^{2}$ band.
\begin{tabular}{cdd}
{\bf Transition}  &  {\bf HWHM (cm$^{-1}$)}  &  {\bf {1$\sigma$}}  \\ \hline
Q$_{3}$(3) & 22.08&0.27\\
R$_{3}$(3) & 26.02&1.36\\
Q$_{6}$(6) & 20.09&0.16\\
R$_{6}$(6) & 21.92&4.07
\end{tabular}
2$^{3}$ band.
\begin{tabular}{cdd}
{\bf Transition}  &  {\bf HWHM (cm$^{-1}$)}  &  {\bf {1$\sigma$}}  \\ \hline
Q$_{1}$(1) & 42.37&0.58\\
Q$_{2}$(2) & 45.30&0.64\\
Q$_{3}$(3) & 45.33&0.30\\
Q$_{3}$(4) & 47.32&0.72\\
Q$_{4}$(4) & 53.33&0.35\\
Q$_{3}$(5) & 48.36&0.23\\
Q$_{5}$(5) & 49.66&0.53\\
Q$_{3}$(6) & 47.94&0.71\\
Q$_{6}$(6) & 46.91&0.33 \\
Q$_{6}$(7) & 48.53&0.22\\
Q$_{7}$(7) & 44.54&0.24\\
Q$_{6}$(8) & 45.41&0.61
\end{tabular}
\end{center}
\label{NH3Widths}
\end{table}

\begin{table}

\caption{Linewidths of the NH$_{2}$D \~A $\leftarrow$ \~X state MODR
transitions.}

\begin{center}
2$^{0}$ band.
\begin{tabular}{cdd}\hline
{\bf Transition}  &  {\bf HWHM (cm$^{-1}$)}  &  {\bf {1$\sigma$}}  \\ \hline
3$_{03}$ $\leftarrow$ 3$_{13}$ & 17.46&0.10\\
4$_{23}$ $\leftarrow$ 3$_{13}$ & 17.46&0.10\\
4$_{14}$ $\leftarrow$ 4$_{04}$ & 17.41&0.10\\
5$_{14}$ $\leftarrow$ 4$_{04}$ & 17.41&0.10\\
5$_{15}$ $\leftarrow$ 5$_{05}$ & 18.57&0.18 \\
6$_{15}$ $\leftarrow$ 5$_{05}$ & 18.57&0.18\\
6$_{06}$ $\leftarrow$ 7$_{16}$ & 22.85&0.46\\
7$_{26}$ $\leftarrow$ 7$_{16}$ & 22.85&0.46\\
7$_{07}$ $\leftarrow$ 8$_{17}$ & 18.97&0.18\\
8$_{27}$ $\leftarrow$ 8$_{17}$ & 18.97&0.18\\
9$_{27}$ $\leftarrow$ 8$_{17}$ & 19.47&1.32
\end{tabular}
2$^{1}$ band.
\begin{tabular}{cdd}\hline
{\bf Transition}  &  {\bf HWHM (cm$^{-1}$)}  &  {\bf {1$\sigma$}}  \\ \hline
3$_{13}$ $\leftarrow$ 3$_{03}$ &  11.74&0.04\\
4$_{13}$ $\leftarrow$ 3$_{03}$ & 12.27&0.26\\
4$_{04}$ $\leftarrow$ 4$_{14}$ & 12.21&0.14\\
5$_{24}$ $\leftarrow$ 4$_{14}$ &  12.21&0.14\\
5$_{05}$ $\leftarrow$ 5$_{15}$ & 12.43&0.09\\
6$_{25}$ $\leftarrow$ 5$_{15}$ & 20.33&1.53\\
6$_{16}$ $\leftarrow$ 7$_{26}$ & 10.35&1.11\\
7$_{16}$ $\leftarrow$ 7$_{26}$ & 13.54&0.14\\
8$_{36}$ $\leftarrow$ 7$_{26}$ & 11.83&0.73\\
7$_{17}$ $\leftarrow$ 8$_{27}$ & 11.64&0.94\\
8$_{17}$ $\leftarrow$ 8$_{27}$ & 13.74&0.08\\
9$_{37}$ $\leftarrow$ 8$_{27}$ & 15.27&0.74
\end{tabular}
2$^{2}$ band.
\begin{tabular}{cdd}\hline
{\bf Transition}  &  {\bf HWHM (cm$^{-1}$)}  &  {\bf {1$\sigma$}}  \\ \hline
3$_{03}$ $\leftarrow$ 3$_{13}$ & 21.30&0.12\\
4$_{23}$ $\leftarrow$ 3$_{13}$ & 21.30&0.12\\
4$_{14}$ $\leftarrow$ 4$_{04}$ & 19.60&0.28\\
5$_{15}$ $\leftarrow$ 5$_{05}$ & 21.85&0.13\\
6$_{15}$ $\leftarrow$ 5$_{05}$ & 21.85&0.13\\
7$_{26}$ $\leftarrow$ 7$_{16}$ & 20.48&0.78\\
7$_{07}$ $\leftarrow$ 8$_{17}$ & 7.83&3.43\\
8$_{27}$ $\leftarrow$ 8$_{17}$ & 21.62&0.40\\
9$_{27}$ $\leftarrow$ 8$_{17}$ & 42.11&7.39\\
\end{tabular}
\end{center}
\label{NH2DWidths}
\end{table}

\begin{table}

\caption{Linewidths of the NHD$_{2}$ \~A $\leftarrow$ \~X state MODR
transitions.}
\begin{center}
2$^{0}$ band.
\begin{tabular}{cdd}\hline
{\bf Transition}  &  {\bf HWHM (cm$^{-1}$)}  &  {\bf {1$\sigma$}}  \\ \hline
2$_{02}$ $\leftarrow$ 2$_{12}$ & 10.24&0.11 \\
3$_{22}$ $\leftarrow$ 2$_{12}$ & 11.12&0.35 \\
3$_{13}$ $\leftarrow$ 3$_{03}$ & 9.41&0.17 \\
4$_{13}$ $\leftarrow$ 3$_{03}$ & 6.58&0.83 \\
4$_{14}$ $\leftarrow$ 5$_{24}$ & 8.18&0.99 \\
5$_{14}$ $\leftarrow$ 5$_{24}$ & 10.30&0.12 \\
6$_{34}$ $\leftarrow$ 5$_{24}$ & 10.09&0.45
\end{tabular}
2$^{1}$ band.
\begin{tabular}{cdd}\hline
{\bf Transition}  &  {\bf HWHM (cm$^{-1}$)}  &  {\bf {1$\sigma$}}  \\ \hline
2$_{12}$ $\leftarrow$ 2$_{02}$ & 6.45&0.04 \\
3$_{12}$ $\leftarrow$ 2$_{02}$ & 5.95&0.14 \\
3$_{03}$ $\leftarrow$ 3$_{13}$ & 6.29&0.05 \\
4$_{23}$ $\leftarrow$ 3$_{13}$ & 6.35&0.28 \\
4$_{04}$ $\leftarrow$ 5$_{14}$ & 6.66&0.21 \\
5$_{24}$ $\leftarrow$ 5$_{14}$ & 6.54&0.03 \\
6$_{24}$ $\leftarrow$ 5$_{14}$ & 7.13&0.10
\end{tabular}
2$^{2}$ band.
\begin{tabular}{cdd}\hline
{\bf Transition}  &  {\bf HWHM (cm$^{-1}$)}  &  {\bf {1$\sigma$}}  \\ \hline
2$_{02}$ $\leftarrow$ 2$_{12}$ & 17.88&0.18 \\
3$_{22}$ $\leftarrow$ 2$_{12}$ & 17.88&0.18 \\
3$_{13}$ $\leftarrow$ 3$_{03}$ & 18.98&0.28 \\
4$_{13}$ $\leftarrow$ 3$_{03}$ & 18.98&0.28 \\
4$_{14}$ $\leftarrow$ 5$_{24}$ & 18.99&0.09 \\
5$_{14}$ $\leftarrow$ 5$_{24}$ & 18.99&0.09 \\
6$_{34}$ $\leftarrow$ 5$_{24}$ & 18.99&0.09
\end{tabular}
\end{center}
\label{NHD2Widths}
\end{table}

\begin{table}

\begin{center}
\caption{Linewidths of the ND$_{3}$ \~A $\leftarrow$ \~X state MODR
transitions.}
2$^{0}$ band.
\begin{tabular}{cdd}\hline
{\bf Transition}  &  {\bf HWHM (cm$^{-1}$)}  &  {\bf {1$\sigma$}}  \\ \hline
Q$_{3}$(3) & 2.241&0.099\\
Q$_{4}$(4) & 1.927&0.051\\
Q$_{3}$(5) & 1.916&0.096\\
Q$_{5}$(5) & 2.024&0.036\\
Q$_{6}$(6) & 1.946&0.075\\
Q$_{6}$(7) & 1.364&0.114\\
Q$_{7}$(7) & 2.085&0.038\\
Q$_{6}$(8) & 1.790&0.056\\
Q$_{7}$(8) & 2.193&0.053\\
Q$_{8}$(8) & 2.075&0.026 \\
Q$_{6}$(9) & 1.581&0.355\\
Q$_{9}$(9) & 1.524&0.062\\
Q$_{9}$(10)& 1.887&0.070
\end{tabular}
2$^{1}$ band.
\begin{tabular}{cdd}\hline
{\bf Transition}  &  {\bf HWHM (cm$^{-1}$)}  &  {\bf {1$\sigma$}}  \\ \hline
Q$_{1}$(1)  &    0.782&0.044 \\
Q$_{2}$(2)  &    0.781&0.030 \\
Q$_{3}$(3)  &    0.664&0.016 \\
R$_{3}$(3)  &    0.742&0.029 \\
Q$_{3}$(4)  &    0.749&0.025 \\
Q$_{4}$(4)  &    0.765&0.008 \\
R$_{4}$(4)  &    0.743&0.059 \\
Q$_{3}$(5)  &    0.832&0.053 \\
Q$_{5}$(5)  &    0.936&0.033 \\
Q$_{3}$(6)  &    1.080&0.053 \\
Q$_{5}$(6)  &    1.050&0.036 \\
Q$_{6}$(6)  &    1.518&0.019 \\
Q$_{6}$(7)  &    0.958&0.017 \\
Q$_{7}$(7)  &    1.049&0.014 \\
Q$_{6}$(8)  &    1.344&0.021 \\
Q$_{7}$(8)  &    1.106&0.050 \\
Q$_{8}$(8)  &    1.039&0.012 \\
Q$_{3}$(9)  &    1.540&0.137 \\
Q$_{6}$(9)  &    1.923&0.127 \\
Q$_{9}$(9)  &    1.281&0.031 \\
Q$_{9}$(10) &    1.611&0.029 \\
Q$_{9}$(11) &    1.849&0.058 \\
Q$_{10}$(11)&    1.593&0.075 \\
Q$_{11}$(12)&    1.598&0.058 \\
Q$_{12}$(14)&    1.534&0.086
\end{tabular}
2$^{2}$ band.
\begin{tabular}{cdd}\hline
{\bf Transition}  &  {\bf HWHM (cm$^{-1}$)}  &  {\bf {1$\sigma$}}  \\ \hline
Q$_{3}$(3) &       7.226&0.010 \\
Q$_{3}$(4) &       11.592&0.328 \\
Q$_{4}$(4) &       7.660&0.198 \\
Q$_{3}$(5) &       9.423&0.463 \\
Q$_{5}$(5) &       9.941&0.172\\
Q$_{6}$(6) &       7.640&0.148\\
Q$_{6}$(7) &       6.814&0.505 \\
Q$_{7}$(8) &       8.789&0.221\\
Q$_{9}$(9) &       8.352&0.704
\end{tabular}
\end{center}
\label{ND3Widths}
\end{table}

\begin{table}
\caption{%
Constants obtained from a fit of the observed rovibronic linewidths to
equation~\protect\ref{Fit_Equation}.%
}
\begin{center}
\begin{tabular}{ccddd}
& {\bf $\nu_{2}$} & {\bf $\Gamma_{o}$} & {\bf $\Gamma_{B}$} & {\bf
$\Gamma_{C}$}\\ \hline
{\bf NH$_{3}$} & 0 & 17.0(4) & 0.006(4) & 0.0028(9) \\
& 1 & 14.6(3) & 0.007(2) & 0.0004(5) \\
& 2 & 22(1) & 0.02(1) & -0.005(2) \\
& 3 & 47(4) & 0.001(4) & -0.001(2) \\ \hline
& & & & \\
{\bf NH$_{2}$D} & 0 & 17.0(9) & & \\
& 1 & 11.5(3) & & \\
& 2 & 21(1) & & \\ \hline
& & & & \\
{\bf NHD$_{2}$} & 0 & 10(1) & & \\
& 1 & 6.4(3) & & \\
& 2 & 17.7(5) & & \\ \hline
& & & & \\
{\bf ND$_{3}$} & 0 & 2.1(1) & -0.0048(29) & -0.0016(11)\\
& 1 & 0.74(5) & 0.0058(31) & 0.0038(15)\\
& 2 & 7.7(8) & 0.065(11) & -0.0021(47)
\end{tabular}
\end{center}
\label{DataFit}
\end{table}

\begin{table}
\caption{Linewidths calculated for 2:1 Fermi coupling between the out-of-plane
bend and the stretching vibration.}
\begin{center}
\begin{tabular}{ccdd}
&$\nu_{2}$ & $\Gamma$(calc) & $\Gamma$(obs)\\
{\bf NH$_{3}$} & & & \\
& 2 & 66.8 & 22 \\
& 3 & 160.4 & 47 \\
& 4 & 244.1 & 75 \\ \hline
& & & \\
{\bf NH$_{2}$D} & & & \\
& 2 & 79.0 & 21 \\
& 3 & 191.5 & 52 \\
& 4 & 313.3 & 78 \\
& 5 & 451.5 & 78 \\ \hline
& & & \\
{\bf NHD$_{2}$} & & & \\
& 2 & 77.7 & 18 \\
& 3 & 207.5 & 36 \\
& 4 & 343.2 & 54 \\
& 5 & 474.6 & 59 \\ \hline
& & & \\
{\bf ND$_{3}$} & & & \\
& 2 & 37.3 & 7.6 \\
& 3 & 98.5 & \\
& 4 & 162.5 & \\
& 5 & 267.2 & 17.5$^{\dag}$
\end{tabular}
$^{\dag}$Reference\protect\cite{Ziegler:1986}
\end{center}
\label{Linewidths}
\end{table}

\begin{thebibliography}{10}

\bibitem{Henck:1990}
{\sc S.~A. Henck},
\newblock {\em The \~A state of Ammonia and Coherence Transfer between
  Rotation-Inversion Transitions in Ammonia},
\newblock PhD thesis, Princeton University, 1990.

\bibitem{Herzberg:1966}
{\sc G.~Herzberg},
\newblock {\em Molecular Spectra and Molecular Structure: Electronic% Spectra
  and Electronic Structure of Polyatonic Molecules}, volume III,
\newblock Van Nostrand Reinhold Company, New York, 1966.

\bibitem{Runau:1977}
{\sc R.~Runau}, {\sc S.~D. Peyerimhoff}, and {\sc R.~J. Buenker},
\newblock {\em J. Mol. Spec.} {\bf 68}, 253 (1977).

\bibitem{McCarty:1987}
{\sc M.~I. McCarthy}, {\sc P.~Rosmus}, {\sc H.~J. Werner}, {\sc P.~Botschwina},
  and {\sc V.~Vaida},
\newblock {\em J. Chem. Phys.} {\bf 86}, 6693 (1987).

\bibitem{Ashfold:1986a}
{\sc M.~N.~R. Ashfold}, {\sc C.~L. Bennett}, and {\sc R.~N. Dixon},
\newblock {\em Faraday Discuss. Chem. Soc.} {\bf 82}, 163 (1986).

\bibitem{Dixon:1988}
{\sc R.~N. Dixon},
\newblock {\em Chem. Phys. Let.} {\bf 147}, 377 (1988).

\bibitem{Rosmus:1987}
{\sc P.~Rosmus}, {\sc P.~Botschwina}, {\sc H.~J. Werner}, {\sc V.~Vaida}, {\sc
  P.~C. Engelking}, and {\sc M.~I. McCarthy},
\newblock {\em J. Chem. Phys.} {\bf 86}, 6677 (1987).

\bibitem{Tang:1988}
{\sc S.~L. Tang} and {\sc D.~G. Imre},
\newblock {\em Chem. Phys. Let.} {\bf 144}, 6 (1988).

\bibitem{Tang:1990}
{\sc S.~L. Tang}, {\sc D.~G. Imre}, and {\sc D.~Tannor},
\newblock {\em J. Chem. Phys.} {\bf 92}, 5919 (1990).

\bibitem{Tang:1991}
{\sc S.~L. Tang}, {\sc E.~V. Abramson}, and {\sc D.~G. Imre},
\newblock {\em J. Phys. Chem.} {\bf 95}, 4969 (1991).

\bibitem{Ashfold:1984}
{\sc M.~N.~R. Ashfold}, {\sc R.~N. Dixon}, and {\sc R.~J. Stickland},
\newblock {\em Chem. Phys.} {\bf 88}, 463 (1984).

\bibitem{Ashfold:1985}
{\sc M.~N.~R. Ashfold}, {\sc C.~L. Bennett}, and {\sc R.~N. Dixon},
\newblock {\em Chem. Phys.} {\bf 93}, 293 (1985).

\bibitem{Ashfold:1990}
{\sc M.~N.~R. Ashfold}, {\sc R.~N. Dixon}, {\sc S.~J. Irving}, {\sc H.~M.
  Koeppe}, {\sc W.~Meier}, {\sc J.~R. Nightingale}, {\sc L.~Schnieder}, and
  {\sc K.~H. Welge},
\newblock {\em Phil. trans. Roy. Soc. London} {\bf 332}, 375 (1990).

\bibitem{Biesner:1988}
{\sc J.~Biesner}, {\sc L.~Schnieder}, {\sc J.~Schmeer}, {\sc G.~Ahlers}, {\sc
  X.~Xie}, {\sc K.~H. Welge}, {\sc M.~N.~R. Ashfold}, and {\sc R.~N. Dixon},
\newblock {\em J. Chem. Phys.} {\bf 88}, 3607 (1988).

\bibitem{Biesner:1989}
{\sc J.~Biesner}, {\sc L.~Schnieder}, {\sc X.~Xie}, {\sc G.~Ahlers}, {\sc K.~H.
  Welge}, {\sc M.~N.~R. Ashfold}, and {\sc R.~N. Dixon},
\newblock {\em J. Chem. Phys.} {\bf 91}, 2901 (1989).

\bibitem{Douglas:1963}
{\sc A.~E. Douglas},
\newblock {\em Dis. Faraday Soc.} {\bf 35}, 158 (1963).

\bibitem{Endo:1990}
{\sc Y.~Endo}, {\sc M.~Iida}, and {\sc Y.~Ohshima},
\newblock {\em Chem. Phys. Let.} {\bf 174}, 401 (1990).

\bibitem{Fuke:1988}
{\sc K.~Fuke}, {\sc H.~Yamada}, {\sc Y.~Yoshida}, and {\sc K.~Kaya},
\newblock {\em J. Chem. Phys.} {\bf 88}, 5238 (1988).

\bibitem{Liefson:1933}
{\sc S.~W. Leifson},
\newblock {\em Astrophysics Journal} {\bf 63}, 73 (1933).

\bibitem{Vaida:1987}
{\sc V.~Vaida}, {\sc M.~I. McCarthy}, {\sc P.~C. Engelking}, {\sc P.~Rosmus},
  {\sc H.~J. Werner}, and {\sc P.~Botschwina},
\newblock {\em J. Chem. Phys.} {\bf 86}, 6669 (1987).

\bibitem{Walsh:1961}
{\sc A.~D. Walsh} and {\sc P.~A. Warshop},
\newblock {\em Trans. Faraday Soc.} {\bf 57}, 345 (1961).

\bibitem{Xie:1986}
{\sc J.~Xie}, {\sc G.~Sh}, {\sc X.~Zhang}, and {\sc C.~Zhang},
\newblock {\em Chem. Phys. Let.} {\bf 124}, 99 (1986).

\bibitem{Ziegler:1984}
{\sc L.~D. Ziegler} and {\sc B.~Hudson},
\newblock {\em J. Phys. Chem.} {\bf 88}, 1110 (1984).

\bibitem{Nakajima:1991}
{\sc A.~Nakajima}, {\sc K.~Fuke}, {\sc K.~Tsukamoto}, {\sc Y.Yoshida}, and {\sc
  K.~Kaya},
\newblock {\em J. Phys. Chem.} {\bf 95}, 571 (1991).

\bibitem{Ziegler:1987}
{\sc L.~D. Ziegler},
\newblock {\em J. Chem. Phys.} {\bf 86}, 1703 (1987).

\bibitem{Ashfold:1991}
{\sc M.~N.~R. Ashfold} and {\sc R.~N. Dixon},
\newblock {\em Chem. Phys. Let.} {\bf 177}, 597 (1991).

\bibitem{Henck:1994a}
{\sc S.~A. Henck}, {\sc M.~A. Mason}, {\sc W.-B. Yan}, and {\sc K.~K. Lehmann},
\newblock Microwave detected, microwave-optical double resonance of NH$_{3}$,
  NH$_{2}$D, NHD$_{2}$, and ND$_{3}$: I. Structure and force field of the \~A
  state.,
\newblock Accepted in J. Chem. Phys.

\bibitem{Dixon:1989}
{\sc R.~N. Dixon},
\newblock {\em Mol. Phys.} {\bf 68}, 263 (1991).

\bibitem{Tannor}
{\sc C.~J. Williams}, {\sc J.~Qiean}, and {\sc D.~J. Tannor},
\newblock {\em J. Chem. Phys.} {\bf 95}, 1721 (1991).

\bibitem{Ziegler:1986}
{\sc L.~D. Ziegler},
\newblock {\em J. Chem. Phys.} {\bf 84}, 6013 (1986).

\end{thebibliography}
\end{document}